\documentclass[twocolumn,showpacs,preprintnumbers,amsmath,amssymb,superscriptaddress,10pt,aps,prb]{revtex4-1}
\usepackage{epsfig,amsopn}
\usepackage{graphicx}
\usepackage{epstopdf}
\usepackage{amsmath,amssymb}
\usepackage{amsthm}
\usepackage{enumerate}
\usepackage{xcolor}
\usepackage[colorlinks=true,linktoc=page,linkcolor=magenta,citecolor=magenta]{hyperref}
\begin{document}
\title{Chiral Luttinger liquids in graphene tuned by irradiation }
\author{Sourav Biswas}
\affiliation{Department of Physics, Indian Institute of Technology - Kanpur, Kanpur 208 016, India}
\author{Tridev Mishra}
\affiliation{Institute for Theoretical Physics, Georg-August-Universit\"at G\"ottingen, Friedrich-Hund-Platz 1, 37077 G\"ottingen, Germany}
\author{Sumathi Rao}
\affiliation{Harish-Chandra Research Institute, HBNI, Chhatnag Road, Jhusi, Allahabad 211 019, India }
\author{Arijit Kundu}
\affiliation{Department of Physics, Indian Institute of Technology - Kanpur, Kanpur 208 016, India}

\begin{abstract}
We show that chiral co-propagating Luttinger liquids can be created and tuned by shining high frequency, circularly polarized light, normal to the layers,  with different polarizations on two  sections of bilayer graphene. By virtue of the broken time-reversal symmetry and the 
resulting mismatch of Chern number, the one-dimensional chiral modes
are localized along the domain wall where the polarization changes. Single layer graphene hosts a single chiral edge mode near each Dirac node, whereas in bilayer graphene, there are two chiral modes near each of the Dirac nodes. These modes, under a high-frequency drive, essentially have a static charge distribution and form a chiral Luttinger liquid under Coulomb interaction, which  can be tuned by means of the driving parameters. We also note that unlike the Luttinger liquids created by electrostatic confinement in bilayer graphene, here there is no back-scattering, and hence our wires along the node are stable to disorder.
\end{abstract}
 
\maketitle

\section{Introduction}

One of the main reasons for the intense interest in bilayer graphene in recent years has been the fact 
that it has a tunable band gap\cite{McCann2006, Castro2007,Oostinga2008,Zhang2009,McCann2012} modulated by an applied gate voltage,
unlike the single layer case\cite{CastroNeto2009} which 
typically require staggered  fields to open up a gap. More recently, it has been realised that it is possible to confine electrons
in gated bilayer structures by applying inhomogeneous electric fields\cite{Martin2008} in such a way that one dimensional states
can be formed at the domain walls separating the two different insulating regions with different gate voltages. These states
are similar to  the zero modes that form at domain walls in polyacetyline\cite{Heeger1988}, superconducting vortices\cite{Semenoff2008}  or 
other solitons in field theories\cite{Jackiw1976}. They are free to move in the direction perpendicular to their confinement
and are  hence  nanowires which  can be tuned by the gate voltages. The effect of electron-electron interactions on such 
wires have also been studied\cite{Killi2010} and it has been demonstrated that these one-dimensional nanowires behave like
strongly interacting Luttinger liquids. Thus bilayer graphene has been shown to be a useful substrate to create and manipulate
strongly interacting one-dimensional quantum wires.

\begin{figure}
\centering
	\includegraphics[width=.4\textwidth]{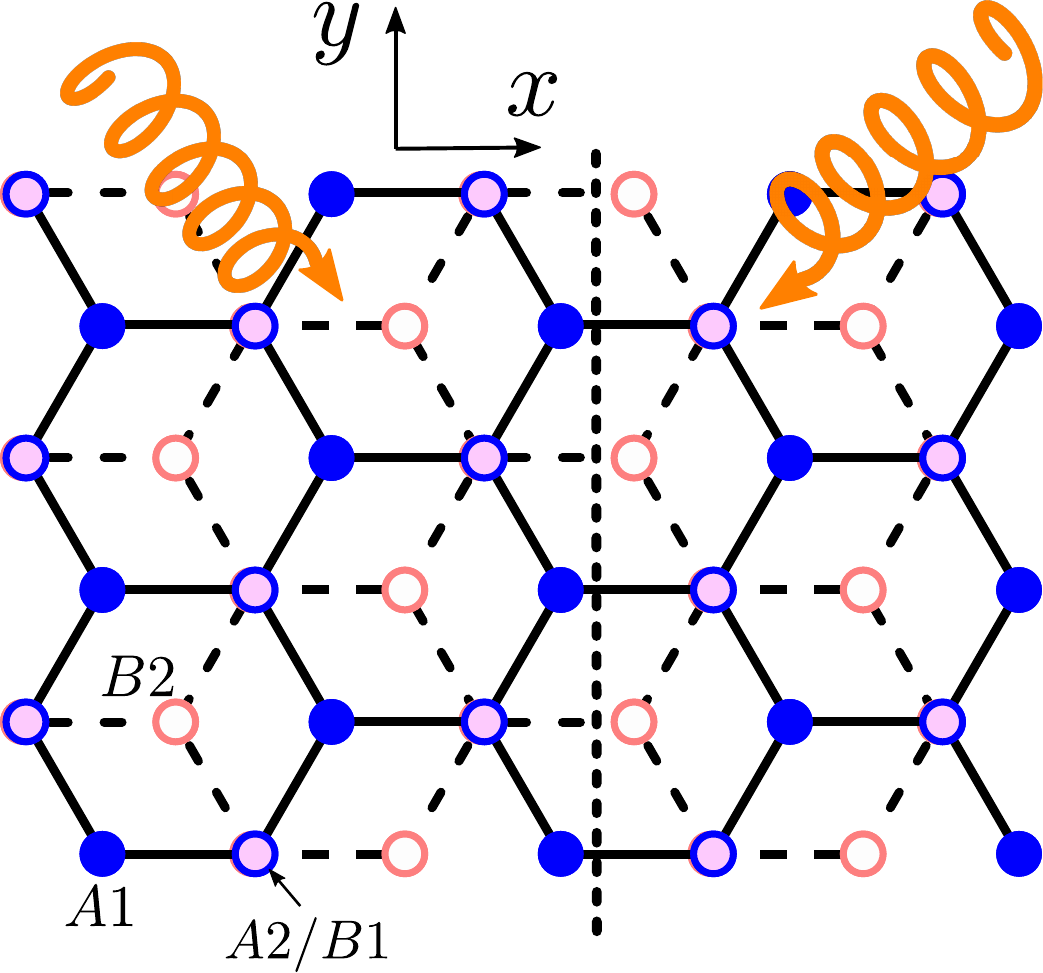}
	\caption{The primary setup of our study: an $A/B$ stacked bi-layer graphene nanoribbon (with a finite width in the $x$-direction) is being irradiated with circularly polarized light of opposite polarizations across a boundary (marked with a dashed line).}\label{fig:setup}
\end{figure}

Floquet engineering\cite{lindner2011,dora2012,rudner2013,goldman2014,farrell2015,titum2016,klinovaja2016,seradjeh2018}
or the generation of new Hamiltonians that are not present in static systems but emerge in driven systems, 
have recently become a very important field of study. In the case of  graphene, it has been realised that the possibility
of tuning the band gap by shining light greatly increases the potential of applications and there has been considerable work\cite{oka2009,kitagawa2011,kundu2014,usaj2014,fkundu2016,mikami2016,mohan2016, mukherjee2017, mishra2018, mishra2015}
on new topological phases obtained by shining light on graphene, as well as bilayer graphene\cite{morell2012,dallago2017,iorsh2017,mohan2018}. Recent experimental observation of anomalous Hall effect in irradiated graphene confirms the Floquet bands and their non-trivial Berry curvature~\cite{graexp}. 
Since shining light
changes the electric field acting on the electrons in graphene or bilayer graphene, a natural question to ask is  whether
it  would be possible to create confinement of electrons using inhomogeneous light instead of an inhomogeneous electric field
using gates.  As we shall see in this paper, the answer to this question can be answered positively.
Moreover, unlike externally applied voltages, shining light breaks the time reversal invariance of the system, and so we
find that the edge states that are created by confinement by light are chiral in nature.

Our main focus in this paper will be  on the chiral edge modes that emerge at the interface  where there is a change in the polarization or phase of the circularly polarized light (CPL) applied perpendicular to the plane of either a single-layer graphene (SLG) or an $A$-$B$ stacked bi-layer graphene (BLG). We shall show that the steady-state edge modes near each of the valleys turn out to be chiral (either both left-handed or both right-handed), since they result from the breaking of time reversal by light.
%As the edge-modes result by breaking the time-reversal symmetry, we argue that, in presence of coulomb interaction, for high frequency drive, the interaction among these modes is essentially time-independent. 
For high frequency driving, these modes are time-independent and in the presence of Coulomb interaction, the interactions
between the modes  are also essentially time-independent.
We shall then show that Coulomb interaction between these chiral modes leads to their mixing,  which can then be rediagonalised
using the standard techniques of bosonisation and Luttinger liquids.  We can then obtain the  power law behaviour of the charge density and spin density  correlation functions and show that the exponents, which should be sensitive to scanning tunneling measurements, can be tuned by changing the amplitude of the impinging radiation. Although we model our system based on single or bi-layer graphene, the qualitative aspects of the resulting chiral Luttinger liquid physics we expect to be model independent and similar treatment can be made for other topological edge-modes of driven symmetry-broken phases.

\section{Irradiation by CPL: High frequency approximation}
We consider the model of bi-layer graphene below and analyze the resulting effective static Hamiltonian in the high-frequency limit. The results of a single-layer of graphene can be recovered in the absence of inter-layer hoppings. The Hamiltonian, for the electrons of each spin, on  bi-layer graphene contains the Hamiltonian of each of the single-layers, $H_{\text{SLG}}$, and a coupling Hamiltonian between the layers, $H_{\text{inter}}$:
\begin{align}
H_{\text{SLG}}&=-t \sum_{\langle i j\rangle,l} a^{\dagger}_{l,i} b_{l,j} + \text{h.c.}
\label{eq:SGH}\\
H_{\text{inter}}&=t_p \sum_{i\in A, j\in B} a^\dagger_{2,i}b_{1,j} + \text{h.c.},\label{eq:inter}
\end{align}
where $l=1,2$ denotes the layer index, and  $a^{\dagger}$ and $b^{\dagger}$ are, respectively, the creation operators for the $A$ and $B$-sublattices of each of the layers. $t$ and $t_p$ are, respectively, the intra and inter-layer hopping amplitudes and we take the estimation $t_p = 0.1t$ with $t=2.7$eV. For each single-layer, if not coupled to a second layer, the electrons follow an effective relativistic dispersion near the two distinct Dirac nodes $K,K'$ with the Fermi velocity $\nu$ given by $\hbar \nu = \frac32ta_0$, where $a_0$ is the lattice constant. For the rest of the paper we set $\hbar=a_0=1$, which serves  as our unit of energy and length, respectively. We consider  Bernal stacking of the two layers, where the inter-layer hopping amplitudes are only between the $B$-sublattice of the top layer and the $A$-sublattice of the bottom layer, as shown in the Fig.~\ref{fig:setup}.

The low energy Hamiltonian, at a single  Dirac node is given by \cite{McCann2006,Martin2008}
\begin{align}
H= \begin{pmatrix}
    0 & \nu \pi^{\dagger} & 0 & 0 \\
    \nu \pi & 0 & t_p & 0\\
    0 & t_p & 0 & \nu \pi^{\dagger} \\
    0 & 0 & \nu \pi & 0
   \end{pmatrix},
\end{align}
written in the basis of the wave-functions $\Psi = (\psi_{A1},\psi_{B1},\psi_{A2},\psi_{B2})$.
The canonical momenta are defined as $\pi = p_x+ip_y, \pi^{\dagger}=p_x-ip_y$,
in terms of the quasimomentum operators $p_x$ and $ p_y$.

\begin{figure*}
	\centering
	\includegraphics[width=.98\textwidth]{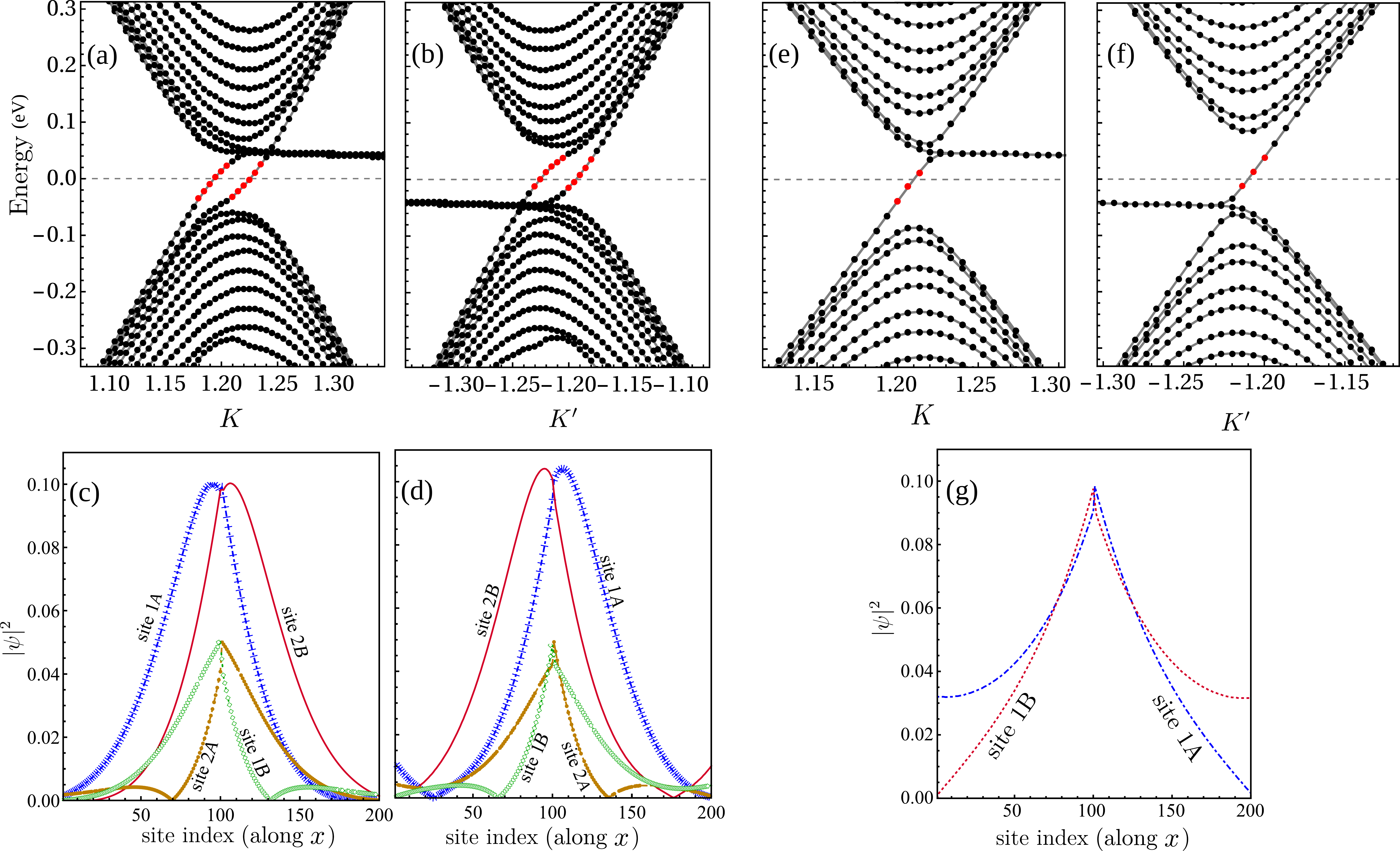}
	\caption{(a) and (b) show the quasi-energy modes (Eq.~\ref{eq:qe}) of the BLG nano-ribbon system with two oppositely polarized irradiation as a function of the momentum along the axis of the ribbon ($k_y$, see Fig.~\ref{fig:setup}), near the $K,K'$ Dirac points (momentum is measured in  units of $1/a_0$, $a_0$ being the lattice spacing of the hexagonal structure). A total of four edge-modes, which run along the boundary between the differently irradiated regions are marked in the figure. For the two edge-modes in (a), their spatial character is depicted in (c) and (d), showing the weight of the steady-state wave-functions in $A$ and $B$ sublattices in layer 1 and 2. The spread of the wave-function depends on the strength and the frequency of the irradiation.
	(e) and (f) shows the corresponding energy dispersion and two  edge modes near $K$, $K'$ points of the SLG nanoribbon. For the edge-mode in (e), the spatial character is depicted in (g), showing the weight of the steady-state wave-functions in $A$ and $B$ sublattices in the single layer 1. A frequency of $\omega/t=30$, an amplitude of  $A_0=0.3$, and 200 sites for each layer (each site contains $A$ and $B$ sub-lattices)  has been used in the numerical simulation.}\label{fig:edge}
\end{figure*}

We now apply high frequency circularly polarized light (CPL)  perpendicular to the plane of the layers.  The  vector potential of the 
radiation is of the  form 
\begin{align}
\mathbf{A}(t) = (A_x\cos(\omega t), A_y\sin(\omega t+ \theta), 0),\label{eq:A}
\end{align}
where $\omega$ is the frequency of
the light and $\theta$ is its polarization angle. To work legitimately in the low energy sector around a
single Dirac node, post the application of radiation, we  need to assume that the amplitudes $A_x$ and $A_y$ are weak enough to allow a linear dispersion approximation to hold.
The Hamiltonian for the bilayer in the presence of such a driving  force 
is given by 
\begin{equation}
 \label{drivenBLG} 
 \mathcal{H}(t) = \begin{pmatrix}
    0 & \nu \tilde{\pi}^{\dagger}(t) & 0 & 0 \\
    \nu \tilde{\pi}(t) & 0 & t_p & 0\\
    0 & t_p & 0 & \nu \tilde{\pi}^{\dagger}(t) \\
    0 & 0 & \nu \tilde{\pi}(t) & 0 
   \end{pmatrix}
\end{equation}
where,
\begin{align}
\label{gaugecoupl}
\tilde{\pi}(t)& = \left(p_x-eA_x\cos(\omega t)\right)+i\left(p_y-eA_y\sin(\omega t+ \theta)\right)\nonumber
  \end{align}
denotes the canonical momentum with the gauge field $\mathbf{A}(t)$ being included in $\mathcal{H}$ in a minimally coupled fashion, with $e$ being
the electronic charge and setting the speed of light $c=1$.
Since we take  the frequency of the light to be very high,  $i.e.$,  much larger than all other scales (such as intra-layer or inter-layer hoppings)  in the problem, 
it is possible to compute an effective time-independent Hamiltonian for the system.  There are several high frequency
approximations\cite{mikami2016,Feldman1984,Mananga2011,Casas2001,Kuwahara2016,Eckardt2015,Bukov2015} that one can use to obtain the static Hamiltonian, all
of which agree to first order in the inverse frequency $1/\omega$. Using such an approximation, one obtains the effective static Hamiltonian, given by
\begin{align}
H_{\rm eff}\approx  \mathcal{H}_0 + \frac{1}{\omega}\displaystyle\sum_{n=1}^{\infty}\frac{[ \mathcal{H}_n, \mathcal{H}_{-n}]}{n} + \mathcal{O}(\omega^{-2})~.
\end{align} 
Here $\mathcal{H}_n$  denote the Fourier coefficients of the periodic time-dependent  Hamiltonian ($ \mathcal{H}(t)$ in Eq.\eqref{drivenBLG} in our case), 
and $[,]$ denotes a commutator bracket. 
Rewriting  the Hamiltonian $\mathcal{H}(t)$ in Eq.\eqref{drivenBLG} 
as 
\[  \mathcal{H}(t)=  \mathcal{H}_0 +  \mathcal{H}_1 e^{i\omega t} +  \mathcal{H}_{-1}e^{-i\omega t}
 \]
we obtain
\[\mathcal{H}_0 = \begin{pmatrix}
    0 & \nu \pi^{\dagger} & 0 & 0 \\
    \nu \pi & 0 & t_p & 0\\
    0 & t_p & 0 & \nu \pi^{\dagger} \\
    0 & 0 & \nu \pi & 0 
   \end{pmatrix}
\]
and the Fourier coefficients $\mathcal{H}_{n=\pm 1}$ as
\begin{widetext}
\[ \mathcal{H}_{\pm 1}= \begin{pmatrix}
                         0 & -\frac{\lambda}{2}(A_x\mp A_ye^{\pm i\theta}) & 0 & 0 \\
                         -\frac{\lambda}{2}(A_x\pm A_ye^{\pm i\theta}) & 0 & 0 & 0 \\
                         0 & 0 & 0 & -\frac{\lambda}{2}(A_x\mp A_ye^{\pm i\theta})\\
                         0 & 0 & -\frac{\lambda}{2}(A_x\pm A_ye^{\pm i\theta})& 0
                        \end{pmatrix}
\]
\end{widetext}
where, $\lambda= \nu e$.  The effective, time-independent Hamiltonian for the BLG system
driven by high-frequency CPL is thus obtained as 
\begin{equation}
 \label{Heff}
 H_{\rm eff}= \begin{pmatrix}
   -\frac{\lambda^2\gamma}{4\omega}\cos\theta & \nu \pi^{\dagger} & 0 & 0 \\
    \nu \pi & \frac{\lambda^2\gamma}{4\omega}\cos\theta & t_p & 0\\
    0 & t_p & -\frac{\lambda^2\gamma}{4\omega}\cos\theta & \nu \pi^{\dagger} \\
    0 & 0 & \nu \pi & \frac{\lambda^2\gamma}{4\omega}\cos\theta
   \end{pmatrix}
\end{equation}
where $\gamma = 4A_xA_y$.
Hence,  the effect of the CPL, in the  
high frequency limit, is essentially to introduce an on-site modulation on the sublattice sites of
the two layers, amounting to a sublattice staggering potential at the $A$ and $B$ sites of both layers.
Thus, we would expect that shining light gives rise to an effective Haldane gap at both the valleys.

In the above effective Hamiltonian, if we set $t_p=0$, we recover the effective Hamiltonian for each of the single layers of the graphene as
\begin{align}
\label{Heffslg}
H^{\rm SLG}_{\rm eff}= \begin{pmatrix}
-\frac{\lambda^2\gamma}{4\omega}\cos\theta & \nu \pi^{\dagger} \\
\nu \pi & \frac{\lambda^2\gamma}{4\omega}\cos\theta,
\end{pmatrix}
\end{align}
which has the form of a two-dimensional Dirac equation with the mass term given by  $m=\frac{\lambda^2\gamma}{4\omega}\cos\theta$. This immediately implies that the sign of the mass term, and thus the Chern number can be modified by changing the sign of $\cos \theta$, near each of the Dirac points. If we have opposite polarization of the irradiation for $x>0$ and $x<0$, then this gives rise to two resulting topological edge-modes at $x=0$, of the same chirality, one each near momenta  $K$ and $K'$.

For the case of bi-layer graphene, our next step is to compute the effective  low energy two band model which describes the physics from the four-band Hamiltonian Eq.~(\ref{Heff}) using a prescription\cite{Manes2007} similar to that which has been used for gated bilayer graphene\cite{Martin2008}.
To do that, we first rewrite the $H_{\rm eff}$ in a  modified 
site basis $(A1,B2,B1,A2)$  where  the effective Hamiltonian
takes the form,
\begin{align}
H_{\rm eff} =& \begin{pmatrix}
   -\frac{\lambda^2\gamma}{4\omega}\cos\theta & 0 & \nu \pi^{\dagger} & 0 \\
    0 & \frac{\lambda^2\gamma}{4\omega}\cos\theta & 0 & \nu \pi\\
    \nu \pi & 0 & \frac{\lambda^2\gamma}{4\omega}\cos\theta & t_p \\
    0 & \nu \pi^{\dagger} & t_p & -\frac{\lambda^2\gamma}{4\omega}\cos\theta
   \end{pmatrix}    \nonumber \\
   \equiv  &\begin{pmatrix} 
                 H_{11} & H_{12}\\
                 H_{21} & H_{22}
                \end{pmatrix} 
\end{align}
where $H_{ij}$ denote the appropriate  $2\times2$ blocks. 
The eigenvalues $\epsilon$ of $H_{\rm eff}$ can then be shown to follow the identity
 \begin{align}    
 {\rm det}(H_{\rm eff}-\epsilon) &= {\rm det}(H_{11}-H_{12}(H_{22}-\epsilon)^{-1}H_{21}-\epsilon) \nonumber \\
& \quad \quad \times{\rm det}(H_{22}-\epsilon).
 \end{align}
 In the low energy regime where $\epsilon\ll t_p$  and
 where $\frac{\lambda^2\gamma}{4\omega}\ll t_p$, since  $H_{22}-\epsilon\simeq H_{22}$,
 we can project the 4-band Hamiltonian   onto the two low energy bands,  given by the Hamiltonian
 \begin{equation}
 \label{eq:lowEnergyHam}
 \begin{split}
  H^L_{\rm eff}&=  H_{11}-H_{12}H^{-1}_{22}H_{21}\\
                &= \begin{pmatrix}
                 - \frac{\lambda^2\gamma}{4\omega}\cos\theta\left(1+ \frac{\nu^2 p^2}{t^2_p}\right)& -\frac{\nu^2\pi^{{\dagger}^2}}{t_p}\\
                 -\frac{\nu^2\pi^{2}}{t_p} & \frac{\lambda^2\gamma}{4\omega}\cos\theta\left(1+ \frac{\nu^2 p^2}{t^2_p}\right)
                \end{pmatrix}
 \end{split}
 \end{equation}
 where $p = (\pi\pi^{\dagger})^{\frac{1}{2}}= \sqrt{p^2_x+p^2_y}$.
Under the approximation where we  drop the second term in the diagonal part of the Hamiltonian, this Hamiltonian becomes
very similar to the Hamiltonian derived for gated bi-layer graphene\cite{Martin2008}. In this approximation, the 
 potential terms on  the diagonal and the momentum dependent terms on the off-diagonal  parts of the Hamiltonian are decoupled. Any 
 position dependence in  this Hamiltonian can now be introduced in the system consistently by simply promoting $p_x$ and $p_y$ 
to their corresponding differential operator representations. This allows the eigenvalue problem to be duplicated by 
a quasi-classical Hamiltonian of the kind derived in Ref.(\onlinecite{Martin2008}), written as
\begin{equation}
\label{eq:qusiclsH}
 \tilde{H}^L_{\rm eff}= -\phi(x)\sigma_z-(p^2_x-p^2_y)\sigma_x-2p_xp_y\sigma_y 
\end{equation}
where $\phi(x) = t^2_p a_0^2/2\nu^2\times  (\lambda^2 \gamma\cos\theta/4\omega)$  in our case and momenta are measured in units
of inverse lattice constant (i.e., $1/a_0$). For the case of a bilayer
strip which is finite in the $x$ direction and translationally invariant in $y$, $p_y$ is a good quantum number whereas $p_x$
is an operator. Hence, we need to solve the  pair of coupled differential equations,
obtained from the eigenvalue problem for the Hamiltonian in eq.\eqref{eq:qusiclsH}, for the potential profile contained in $\phi(x)$. When $\phi(x)$ has a step profile, changing sign across a boundary, topological zero-energy modes are localized at the kink or the domain wall.  In our case, we introduce a position dependence in the polarization angle of the irradiation 
and study the domain wall obtained when we have light of two different polarizations for $x<0$ and $x>0$. 

In fact, the Chern number associated with the ground-state of Eq.~(\ref{eq:qusiclsH}) is given by sgn($\phi$), which is easily seen as following. We rewrite the effective Hamiltonian, Eq.~(\ref{eq:qusiclsH}), as
\begin{align}
\tilde{H}_{\text{eff}}^{L} &\equiv
\begin{bmatrix}
-\phi & k^2 \text{e}^{2i\theta} \\
k^2 \text{e}^{-2i\theta} & \phi \\
\end{bmatrix},
\end{align}
with $k = \sqrt{k^2_y + k^2_x}$ and $\theta = \tan^{-1}(\frac{k_x}{k_y})$. Diagonalization gives the the wave-functions 
\begin{align}
\Psi_{\pm} =
\begin{bmatrix}
\text{e}^{i \theta} \frac{k^2}{\sqrt{k^4 + (\phi \pm \sqrt{\phi^2 + k^4})^2}} \\
\text{e}^{-i \theta} \frac{\phi \pm \sqrt{\phi^2 + k^4}}{\sqrt{k^4 + (\phi \pm \sqrt{\phi^2 + k^4})^2}} \\
\end{bmatrix}
\end{align}
with energies $\epsilon_{\pm} = \pm \sqrt{ \phi^2 + k^4 }$. For the lower energy band, the Chern number is then readily given by 
\begin{align}
\frac{i}{2\pi } \int d^2k [ \langle \partial_{k_x} \Psi_{-} | \partial_{k_y} \Psi_{-} \rangle - \langle \partial_{k_y} \Psi_{-} | \partial_{k_x} \Psi_{-} \rangle  ]=\text{sgn}(\phi).
\end{align}
This implies a change of Chern number of $\Delta C$ = 2 across the boundary of regions with different signs of $\phi$, i.e, in the case where the  two sides are driven with different polarizations $\phi = +\pi$ and $\phi = -\pi$. 

All of this is at a single Dirac point. At the other Dirac point, the operators $\pi$ and $\pi^\dagger$ are interchanged.  It is easy to check that this only leads
to a change in sign in the effective potential term $\phi(x)$ in Eq.~(\ref{eq:qusiclsH}). So essentially, the operators at the $K$ and $K'$ points
are related by the symmetry $\phi(x) \rightarrow -\phi(x),~p_y \rightarrow -p_y$.  This, in turn, comes from the fact that the effect of radiation essentially acts
like a time-reversal symmetry breaking staggered potential and so, unlike in the gated bi-layer graphene system,  the edge states at both the $K$ and the $K'$ valleys
have the same chirality (decided by the chirality of the circular polarization of the impinging light).

\section{Edge mode steady states at the domain wall}
In the previous section, we discussed how the polarized irradiation induces topological gaps at the $K,K'$ Dirac points. If the regions irradiated by 
opposite polarizations are separate, as shown in Fig.~\ref{fig:setup}, one expects edge modes at the interface. If the two sides are irradiated by right and left circularly polarized light (i.e, $\theta = 0,\pi$ on the two sides with $A_x=A_y=A_0$), the net change of Chern number across the interface is two (four) for single-layer (bi-layer) graphene and accordingly one expects two (four) chiral modes to run along the interface. We study the edge-modes through a tight-binding simulation, by incorporating the vector-potential Eq.~(\ref{eq:A}) in the Hamiltonian Eq.~(\ref{eq:SGH}) by Peierls substitution, neglecting the phase difference in $\vec{A}(t)$ between the two layers, which is justified as $l_{\text{int}}\omega/c\ll 1$, where $c$ is the velocity of light and  $l_{\text{int}}$ is the inter-layer distance. The steady states of the time-periodic Hamiltonian, $H(t)$, after the Peierls substitution can be found using the  Floquet theorem, which states that the eigen-states will be of the form
\begin{align}
\psi_{\alpha}(t) = e^{-i\epsilon_{\alpha}t}u_{\alpha}(t),
\end{align}
where $\epsilon_{\alpha}$s are the `quasienergies' and $u_{\alpha}(t)$ are periodic functions, called the Floquet-states, both of which can be found from the eigen-system problem
\begin{align}\label{eq:qe}
(i\partial_t - H(t))u_{\alpha}(t) = \epsilon_{\alpha} u_{\alpha}(t).
\end{align}
For a high-frequency drive, the quasi-energy spectrum of a nano-ribbon, as depicted in Fig.~\ref{fig:setup}, is shown in Fig.~\ref{fig:edge}, highlighting the edge-modes. The edge-modes appear at each $K/K'$ points with a slightly different Fermi velocity $v_1$ and $v_2$.  (The Appendix has  a discussion of how higher orders in the high frequency expansion lead to the fact that $v_1\neq v_2$).

%------------------------------------------------
\begin{figure}
	\centering
	\includegraphics[width=.495\textwidth]{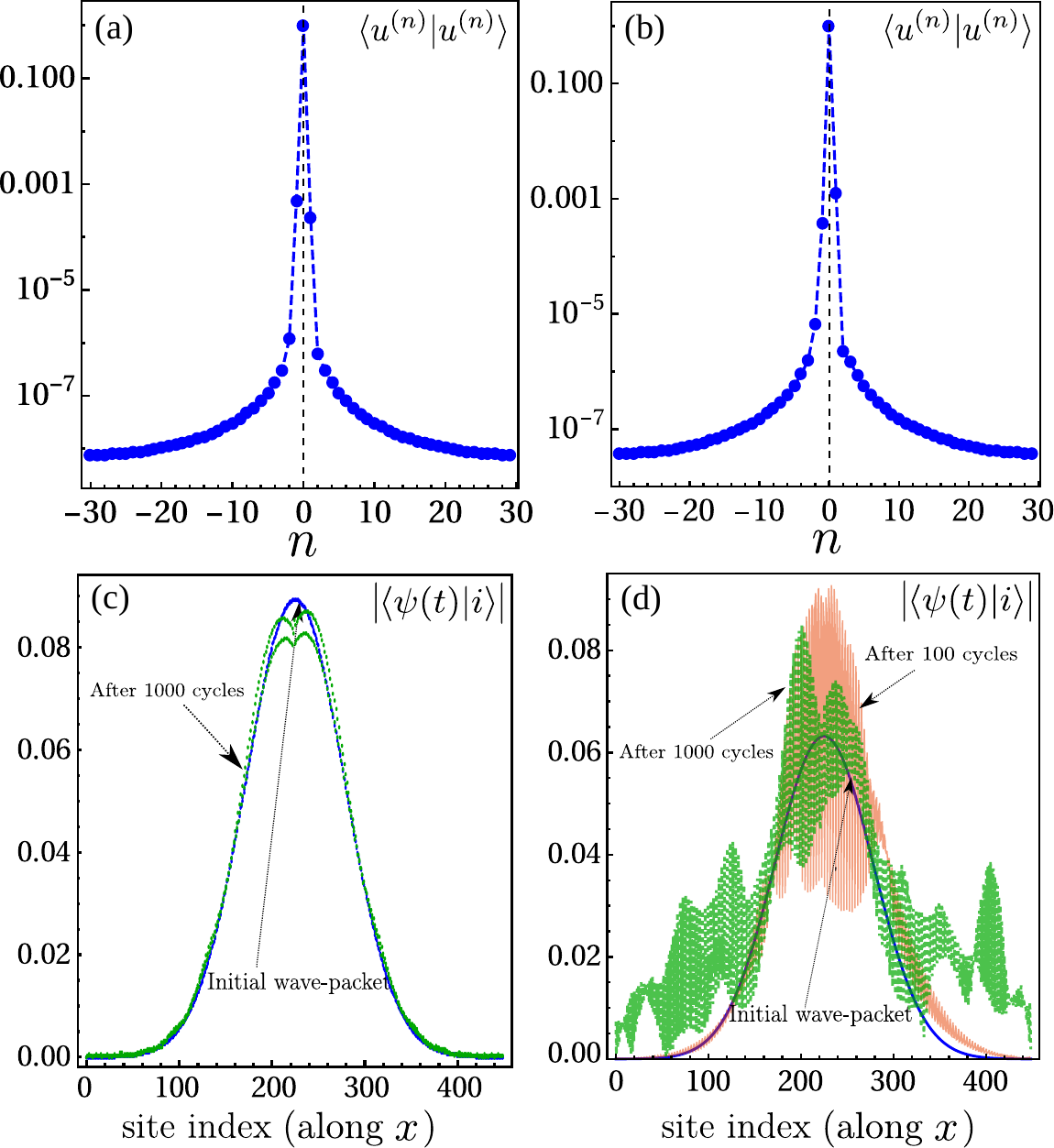}
	\caption{(a) and (b): For one of the edge-modes, the weight of the discrete Fourier components of the Floquet states is plotted (in log-scale), showing that the $n=0$ components are several orders of magnitude larger than the other components, giving rise to  essentially time-independent Floquet modes and allowing for  an effective time-independent interaction among them. (a) corresponds to the edge-mode in SLG and (b) corresponds to one of the edge-modes in BLG. (c) and (d): we show the dynamics of a Gaussian wave-packet (which is not an eigen-state) introduced at  the interface of the system with two different polarizations.  Even after thousands of cycles, the wave-packet remains essentially confined to the edge. In (c) we demonstrate the  effect in BLG, where we show an initial Gaussian wave-packet and its modification after 100 as well as 1000 cycles of  the drive. In (d) we depict  the dynamics at the edge-mode in SLG, where we show an initial Gaussian wave-packet and its modification after 1000 cycles of the drive. For both (c) and (d) 
	the parameters used were $\omega/t=30$ and $A_0=0.3$ }\label{fig:just}
\end{figure}
%------------------------------------------------

Before we proceed to analyze the properties of the topological edge-modes under Coulomb interactions, which we introduce perturbatively, a justification of the use of the non-interacting Floquet analysis is due. A number of recent works\cite{heating1,heating2,heating3,heating4} argue that, for a rapidly driven closed interacting system, the heating time ($\tau_h$) scale is exponentially large in the driving frequency and in the intermediate time the system's dynamics is governed by an effective Hamiltonian, which one may obtain from a high-frequency approximation, such as the van Vleck expansion used in our system. If our system is weakly connected to an environment, giving rise to a relaxation time scale $\tau_l$, then as long as $\tau_l\ll \tau_h$, one expects the system to not be heated. Further, with a  high-frequency drive, even when there is a gap-opening due to the breaking of a  symmetry, such as in our case, one expects no population inversion~\cite{heating5, heating6}. This allows us to consider the occupations to
 have the standard Fermi-Dirac distribution.
% be dictated with the same Fermi-energy.
We further consider that the bulk system, which has Dirac dispersion near the Fermi-energy, is essentially non-interacting and that the Coulomb interaction is only important in the edge-modes. This assumption, strictly speaking, needs to be further justified, and  can be argued as follows. First, since the edge-modes are  topological, they are expected to be robust against weak interactions in the bulk. Second, since the edge
modes are one-dimensional, the effect of Coulomb interaction among them can not be neglected.

Assuming that the preceding approximations hold, we further proceed to make an another argument to justify the application of  Luttinger liquid theory -
namely, we argue that the interaction among the edge-modes are also effectively time-independent. In the presence of perturbative interactions, the effective interaction elements among modes with similar quasi-energies can  be written as
\begin{align}
\langle\psi_{\alpha}(t)|\hat{V}_{\text{int}}|\psi_{\alpha'}(t)\rangle = \langle u_{\alpha}(t)|\hat{V}_{\text{int}}|u_{\alpha'}(t)\rangle,
\end{align}	
where we used the fact that $\epsilon_{\alpha}=\epsilon_{\alpha'}$.  Expanding  in Fourier components, $|u_\alpha(t)\rangle = \sum_n e^{-in\omega t}|u_\alpha^{(n)}\rangle$, the right hand side of the above equation  can be written as 
\begin{align}
=\sum_{n,m} e^{i(n-m)\omega t}\langle u^{(n)}_{\alpha}|\hat{V}_{\text{int}}|u_{\alpha'}^{(m)}\rangle.
\end{align}
We show in Fig.~\ref{fig:just} that for the edge-modes, $|u_{\alpha}^{(0)}\rangle$ is dominant, and other Fourier components can be neglected (i.e, the time-dependence of $\psi_{\alpha}(t)$ is governed by only the dynamical phase factor). This essentially results from the fact that the gap opening (and thus the resulting edge-modes) at the $K,K'$ points takes place even for an infinitesimal driving amplitude without any relevant change of occupation number~\cite{heating5, heating6}. This allows us to simplify the interaction matrix elements among the edge modes to be effectively time-independent.

%------------------------------------------------
\begin{figure}
	\centering
	\includegraphics[width=.4\textwidth]{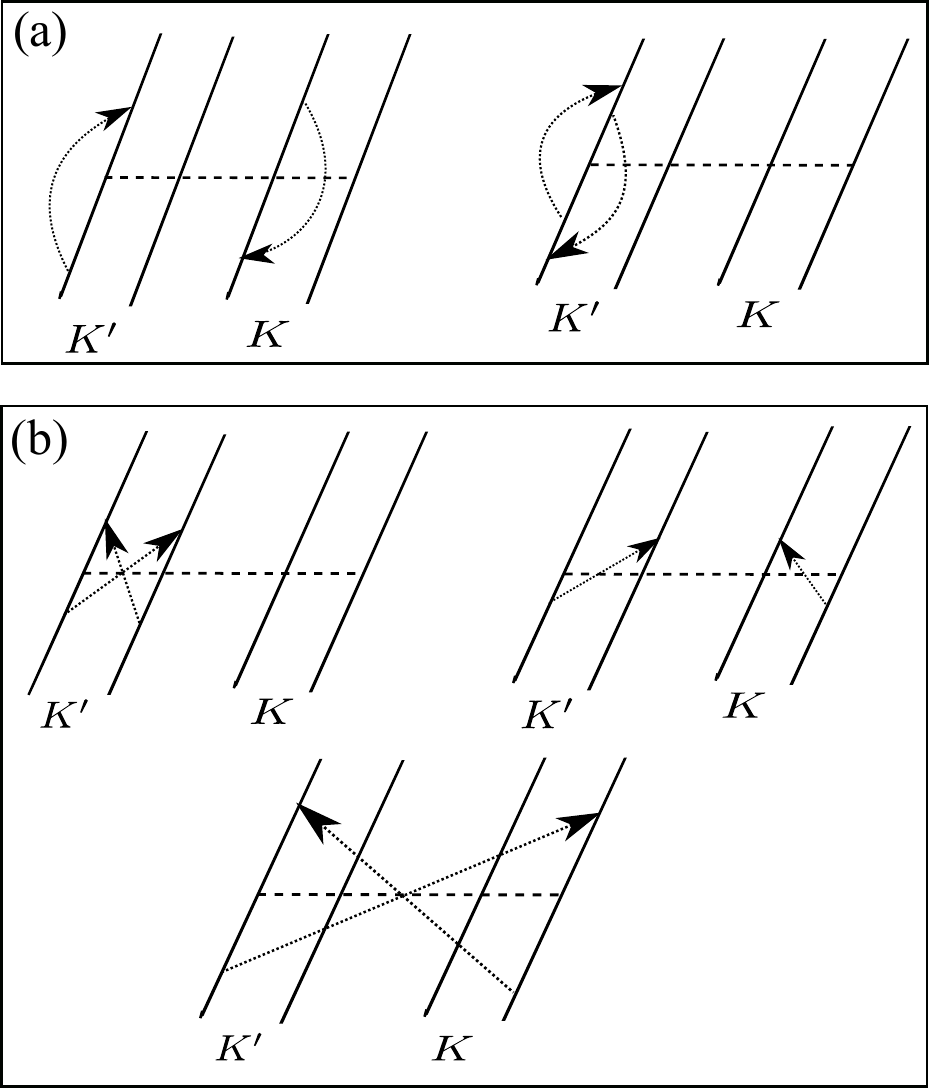}
	\caption{Various types of scattering processes allowed by the interaction Hamiltonian among the edge-modes in BLG. (a) shows the processes which are  of the density-density type (class-I), whereas (b) shows all the  inter-mode scattering processes (class-II). The processes in (b) are sub-dominant by several orders of magnitude and are neglected in our analysis. Some of the processes are naturally not possible in case of  SLG.}\label{fig:process}
\end{figure}
%------------------------------------------------

These edge-states are also expected from the effective static Hamiltonian obtained by the  high-frequency approximation in the last section.  Since there
exists a difference in Chern number between regions of different sign of $\phi$ (i.e, of opposite polarization), it is clear that the regions are topological
and that the boundary must have edge states.  The simplest possibility ( in the absence of further counter-propagating states) is to have as many chiral edge states as dictated by the change in Chern number.  As the effective static Hamiltonian can only predict dynamics in stroboscopic times~\cite{Eckardt2015}, the states may still have significant transverse dynamics. We check this in Fig.~\ref{fig:just}, where, we examine the transverse dispersion of a wave-packet, initially  introduced  at the domain-wall of the two topologically distinct regions, when driven by the time-periodic Hamiltonian. The results show that, for a few hundred cycles, the wave-packet can be considered to be confined at the interface of the two regions, although for BLG (unlike for SLG), the wave packet starts spreading for about a 1000 cycles. In contrast, if the dynamics had been strictly  driven by a time-independent effective Hamiltonian, as was derived in the last section, we would have found  that the wave-packet  would have remained confined to the interface  for a much longer time. This hints at a non-vanishing transverse velocity and a  limit to  the time-scale for the validity of the effective one-dimensional nature of the low-energy excitations.

It should also be noted that if the original wave-packet had been  prepared in the state of $|u_{\alpha}(t)\rangle$ (or, equivalently, $\approx |u_{\alpha}^{(0)}\rangle$), the wave-packet would have been confined to  the boundary for an infinite time, as the Floquet-states are eigenstates of the Hamiltonian. This, however,  would require  careful initial state-preparation, which may  not be an easy task.

\section{Luttinger liquid analysis}
As argued in the previous section, the steady-state edge-modes in this system are essentially time-independent, allowing us to consider an effective time-independent interaction Hamiltonian, which is crucial for our use of  Luttinger liquid theory~\cite{bukov2012}. Keeping this in mind, we write the interaction among the (effectively time-independent) edge-states of the time-periodic Hamiltonian as
\begin{align}
&	H_{int} = \frac12\int d\vec{r}d\vec{r'} \hat{\rho}(\vec{r})V(\vec{r}-\vec{r'})\hat{\rho}(\vec{r'}) ~.
\end{align} 
Here we assume that $\hat{\rho}(r)$ has a trivial time-dependence, which we justify as follows. We write the field operator of the driven system as $\Psi^{\dagger}(\vec{r},t) = \sum_{\alpha}\langle \vec{r}| u_{\alpha}(t)\rangle a_{\alpha}^{\dagger} = \sum_{\alpha,n} e^{i(\epsilon_{\alpha}+n\omega)t} \langle \vec{r}| u^{(n)}_{\alpha}\rangle a_{\alpha}^{\dagger}$, where $|u_{\alpha}(t)\rangle = \sum_{n}e^{-in\omega t}|u^{(n)}_{\alpha}\rangle$ are the Floquet states as defined earlier as well. Then,
\begin{widetext}
\begin{align}
\hat{\rho}(\vec{r},t) &= \Psi^{\dagger}(\vec{r},t)\Psi(\vec{r},t) = \sum_{\alpha,\beta}\sum_{n,m}e^{i(\epsilon_{\alpha}-\epsilon_{\beta})t}\left[e^{i(n-m)\omega t} \langle u^{(m)}_{\beta}| \vec{r}\rangle\langle \vec{r}| u^{(n)}_{\alpha}\rangle  \right] a^{\dagger}_{\alpha}a_{\beta}\nonumber\\
& = \sum_{\alpha,\beta}e^{i(\epsilon_{\alpha}-\epsilon_{\beta})t}\left[\sum_{n}\langle u^{(n)}_{\beta}| \vec{r}\rangle\langle \vec{r}| u^{(n)}_{\alpha}\rangle  \right] a^{\dagger}_{\alpha}a_{\beta} + \sum_{\alpha,\beta}e^{i(\epsilon_{\alpha}-\epsilon_{\beta})t}\left[\sum_{n,m \neq n}e^{i(n-m)\omega t}\langle u^{(m)}_{\beta}| \vec{r}\rangle\langle \vec{r}| u^{(n)}_{\alpha}\rangle  \right] a^{\dagger}_{\alpha}a_{\beta}.\nonumber
\end{align}
\end{widetext}
So, if $\langle \vec{r}|u^{(0)}\rangle \gg \langle \vec{r}|u^{(n\neq 0)}\rangle $ is satisfied, then generally the non-trivial time dependence of the above equation (second part) can be dropped. As mentioned before, we  have verifed this in Figs.~\ref{fig:just}(a) and  \ref{fig:just}(b). Further, we also studied wave-packet dynamics
of a Gaussian wave-packet  in Figs.~\ref{fig:just}(c) and \ref{fig:just}(d)  where we obtained a  time-scale below  which any transverse motion due to the dynamics of the edge-states can be neglected and a Luttinger liquid formalism based on the  time-independent density operator can be justified.

Within the effective time-independent description, at low-energy, we have one chiral mode at each of the Dirac points $K,K'$ for each spin ($\sigma$) for the SLG and we have two chiral modes at each of the Dirac points $K,K'$ for each spin ($\sigma$) for the BLG. We index these modes  by $\alpha = K,K'$ for the SLG and $\alpha = 1_K, 2_K, 1_{K'}, 2_{K'}$ for  the BLG. Assuming translation invariance along the  $y$ direction, we can further write the field operator for each mode, taking into account only the modes near the Fermi energy, with $ \vec{R} = (x,y)$  as
\begin{align}
\hat{\Psi}_{\sigma}(\vec{R}) =\hat{\Psi}_{\sigma}(x,y)=\sum_{\alpha}\phi_{\alpha}(x) e^{i k^{\alpha}_{F} y} \hat{\xi}_{\alpha\sigma}(y) ~,
\end{align} 
where $\hat{\xi}_{\alpha\sigma}(y)$ is a slowly varying function along $y$. There are a total of eight modes (two modes near each of the Dirac points $K,K'$, which are degenerate in spins). The interaction Hamiltonian can then be expressed as
\begin{align}
\tilde{H}_{int} =& \frac{1}{2L_y} \sum_{\sigma \sigma'yy' } \int dy dy' ~ \hat{\Psi}^{\dagger}_{\sigma}(\vec{R})\hat{\Psi}^{\dagger}_{\sigma'}(\vec{R'}) V(| \vec{R}-\vec{R'} |)\nonumber \\
& ~~~~~~~~~~~~~~~~~~~~~~~~~~~~~~~~~~ \times \hat{\Psi}_{\sigma'}(\vec{R'})\hat{\Psi}_{\sigma}(\vec{R}) ~,\nonumber \\
& \equiv  \frac{1}{2} \sum_{\sigma \sigma' y y' \{\alpha \} } \int dy dy'\ \ h,
\end{align} 
where the integrand  can be written as 
\begin{align}
h &=  \frac{e^{iy\Delta k-i\bar{y}(k^{\gamma}_{F}-k^{\beta}_{F})} }{\sqrt{{\bar{y}}^2 + (x-x')^2}} \phi^{*}_{\alpha}(x) \phi^{*}_{\beta}(x') \phi_{\gamma}(x') \phi_{\delta}(x)\nonumber\\
& \quad\quad ~~~~~~~\hat{\xi}_{\alpha\sigma }^{\dagger}(y) \hat{\xi}_{\beta\sigma'}^{\dagger}(y-\bar{y})
\hat{\xi}_{\gamma\sigma'}(y-\bar{y}) \hat{\xi}_ {\delta\sigma}(y),\label{eq:h}
\end{align} 
where $\bar{y} = y-y'$ and considering the functions $\hat{\xi}$ to be slowly varying, momentum conservation requires  $\Delta k=0$.
 Note that we have assumed that the potential $V(|{\bf R} - {\bf R'}|)$ is of the form $e^2/\sqrt{{{\bar{y}}^2 + (x-x')^2}}$.

%------------------------------------------------
\begin{figure*}
	\centering
	\includegraphics[width=.98\textwidth]{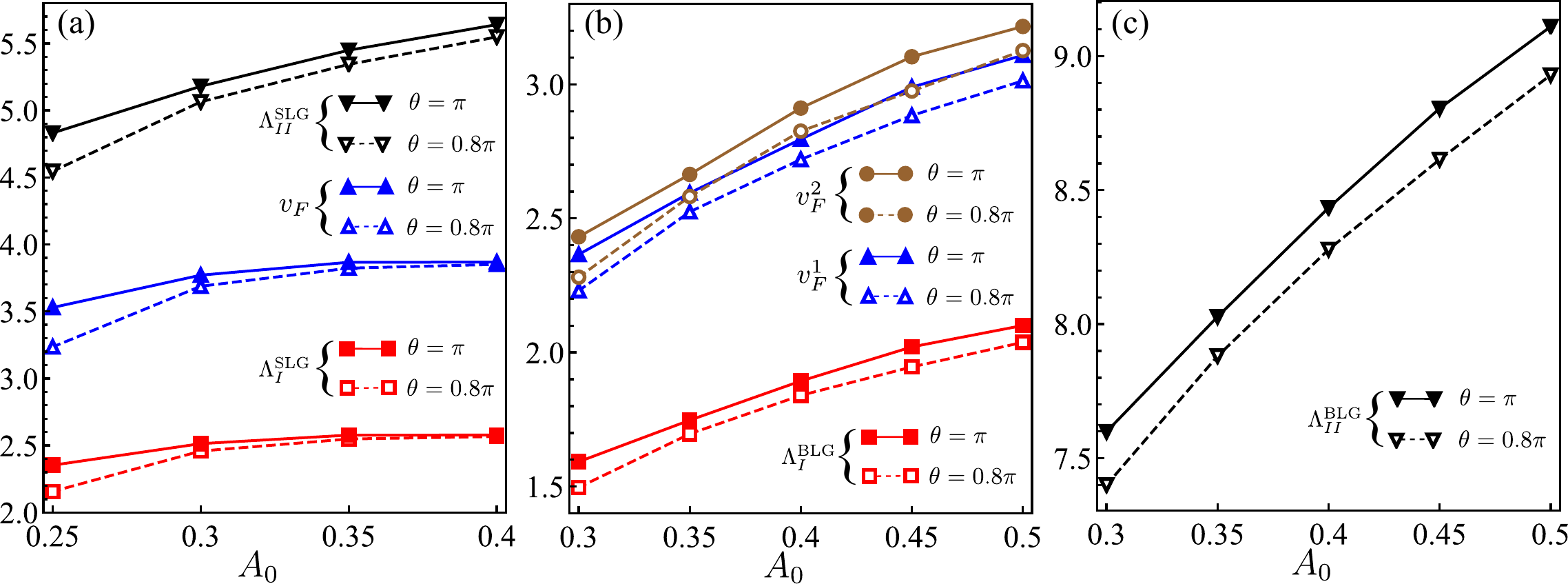}
	\caption{The renormalized velocities of the edge-modes under Coulomb interaction as a function of the driving amplitude $A_0$ and the difference ($\theta$) of the polarization angle of the irradiation  between the left and right halves of the nano-ribbon as shown in Fig.\ref{fig:setup}.  $\theta=\pi$ represents the maximum difference - the case where the two drives are right and left circularly polarized. In (a), we show the two renormalized velocities $\Lambda^{\rm SLG}_{I}$ and $\Lambda^{\rm SLG}_{II}$ for the edge-modes of the SLG setup. Here  $v_F$ is the velocity of the two original modes. Similarly in (b) and (c) we show the result for the BLG system, where only the two modes of the $(+)$ sector are renormalized (see the main text), with renormalized velocities $\Lambda_{I}^{\rm BLG}$ and $\Lambda_{II}^{\rm BLG}$. The velocities are measured in units of $\nu=3ta_0/2\hbar$.}\label{fig:v1v2}
\end{figure*}
%------------------------------------------------

Broadly speaking, the possible scattering processes can be divided into two classes, shown as class I and class II in Fig.~\ref{fig:process}.
Computing the bare scattering amplitudes of all the processes,  we find that  density-density type interactions (class I)  are the dominant ones by several orders
of magnitude and  hence, we keep only such processes. Furthermore, since all scatterings in class I take place in the same mode, we may take $k_F^\gamma = k_F^\beta$. 
 In this case,
the form of $h$ can be written as 
\begin{align}
h  & =\frac{1}{\sqrt{\bar{y}^2 + (x-x')^2}} \phi^{*}_{\alpha}(x) \phi^{*}_{\beta}(x') \phi_{\beta}(x') \phi_{\alpha}(x)\nonumber \\
& \quad\quad ~~~~~~~\hat{\xi}_{\alpha\sigma }^{\dagger}(y) \hat{\xi}_{\beta\sigma'}^{\dagger}(y-\bar{y})
\hat{\xi}_{\beta \sigma'}(y-\bar{y}) \hat{\xi}_ {\alpha \sigma}(y).
\end{align} 
We may now write the one-dimensional form of the interaction Hamiltonian in terms of the  standard two-body scattering amplitudes
 $V_{\alpha\alpha'\alpha'\alpha}$ defined below as 
\begin{align}
\tilde{H}_{int} &\approx \frac{\alpha_g}{2} \int dy \sum_{\alpha \alpha'} V_{\alpha \alpha' \alpha' \alpha}\nonumber\\
&~~~~~~~~\times \sum_{\sigma \sigma'}
\hat{\xi}_{\alpha\sigma }^{\dagger}(y) \hat{\xi}_{\alpha'\sigma'}^{\dagger}(y)
\hat{\xi}_{\alpha' \sigma'}(y) \hat{\xi}_ {\alpha \sigma}(y),
\end{align}
where the scattering amplitudes are given by 
\begin{align}
&V_{\alpha \alpha' \alpha' \alpha} = \sum_{x x' \bar{x} \{ \alpha \} }  \frac{1}{\sqrt{\bar{y}^2 + (x-x')^2}}\nonumber\\
& ~~~~~~~~~~~~~~~~~~~~ \times\phi^{*}_{\alpha}(x) \phi^{*}_{\alpha '}(x') \phi_{\alpha'}(x') \phi_{\alpha}(x) .
\end{align} 
An effective fine-structure constant $\alpha_g\approx 2$ for single-layer graphene is used,  and we write all velocities in terms of the velocity of electrons in graphene. For bosonization we adopt the following notations: $\hat{\xi}_{\alpha\sigma}(y) \sim e^{-i 2\sqrt{\pi} \Phi_{\alpha \sigma}(y)}; ~	 \rho_{\alpha \sigma}(y) = \hat{\xi}^{\dagger}_{\alpha\sigma}(y) \hat{\xi}_{\alpha\sigma}(y)  = -\frac{1}{\sqrt{\pi}} \nabla \Phi_{\alpha \sigma}.$ 
 Here $\Phi_{\alpha\sigma}(y)$  is the bosonic field operator. Note  that we have not included Klein factors, as they can be set to unity in the density and  they do not affect the correlation functions that we compute in this paper. 
In this notation, the density-density interaction becomes 
\begin{equation}
\begin{aligned}
\rho_{\alpha \sigma} \rho_{\alpha' \sigma'} 
=&	~\frac{1}{\pi} \nabla\Phi_{\alpha \sigma}(y) \nabla\Phi_{\alpha' \sigma'}(y).
\end{aligned}
\end{equation} 
Writing $\tilde{H} = \tilde{H}_{o} + \tilde{H}_{int} = \int dx [ H_{0} + H_{int} ]  = \int dx H $, the Hamiltonian can be written in the bosonic language
as 
\begin{align}
H
= H_{0} +H_{int}
=& \sum_{ \alpha \sigma} v^{\alpha}_{F} (\nabla\Phi_{\alpha \sigma})^{2}\nonumber\\
 +& \frac{1}{2 \pi} \sum_{\alpha \alpha' }  V_{\alpha \alpha' \alpha' \alpha}  \sum_{\sigma \sigma' }  \nabla\Phi_{\alpha \sigma} \nabla\Phi_{\alpha' \sigma'}.
\end{align}
where the sums over the $\alpha$ and $\sigma$ indices include all the modes at both Dirac points.
We further introduce bosons  corresponding to different charge, spin sectors for different channels as follows -
\begin{align}
&	\Phi_{\alpha c} = \frac{\Phi_{\alpha \uparrow} + \Phi_{\alpha \downarrow}}{\sqrt{2}}; ~\Phi_{\alpha s} = \frac{\Phi_{\alpha \uparrow} - \Phi_{\alpha \downarrow}}{\sqrt{2}},
\end{align}
where	$\alpha =  K, K'$ (SLG) or $1_K, 2_K, 1_{K'}, 2_{K'}$ (BLG) are different channels, $c$ denotes the charge sector and $s$ denotes the spin~sector. Simplifying we get,
\begin{equation}
\begin{aligned}
\sum_{\sigma \sigma' } \nabla \Phi_{\alpha \sigma} \nabla \Phi_{\alpha' \sigma'}&	=
2 \nabla  \Phi_{\alpha c} \nabla \Phi_{\alpha' c}.
\end{aligned}
\end{equation} 
So the Hamiltonian can be written in terms of the charge and spin bosons as
\begin{align}
H_{0} &= \sum_{\alpha} v^{\alpha}_F [ (\nabla \Phi_{\alpha c})^{2} + (\nabla \Phi_{\alpha s})^{2} ], \\
H_{int} &= \frac{1}{\pi} \sum_{\alpha \alpha' } V_{\alpha \alpha' \alpha' \alpha}  \nabla \Phi_{\alpha c} \nabla \Phi_{\alpha' c},
\end{align}
 where we note that the Coulomb interaction term modifies  only the charge sector.  Thus, in the absence of scatterings involving spin,
the $SU(2)$ spin symmetry is intact and  the spin sector is not expected to be renormalized. 

\subsection{Single-layer graphene}
The  Hamiltonian for the charge sector of SLG can be written as 
\begin{align}
H_{c}^{\rm SLG} =    & 
\begin{bmatrix}
\nabla \Phi_{K c} & \nabla \Phi_{K' c}
\end{bmatrix}
R^{\rm SLG}
\begin{bmatrix} \nonumber
\nabla \Phi_{K c} \\
\nabla \Phi_{K' c}
\end{bmatrix} , \\ 
R^{\rm SLG} =& 
\begin{bmatrix}
v_F^{K}+\frac{1}{\pi}V_{A} & \frac{1}{\pi}V_{B} \\
\frac{1}{\pi}V_{B} & v_F^{K'} + \frac{1}{\pi}V_{A},
\end{bmatrix} 
\end{align}
where we have explicitly used the form of the resulting scattering matrix.  Here $V_A = V_{\alpha\alpha\alpha\alpha}$ and $V_B = V_{\alpha\alpha'\alpha'\alpha}$ ($\alpha\neq\alpha'$). 
This sector can then be diagonalized using the canonical transformation
\begin{align}
\Phi_{K c} = \cos\Theta_s \tilde{\Phi}_{1 c} + \sin\Theta_s \tilde{\Phi}_{2 c} ,\\ \nonumber
\Phi_{K' c} = -\sin\Theta_s \tilde{\Phi}_{1 c} + \cos\Theta_s \tilde{\Phi}_{2 c},	
\end{align} 
with $\tan(2\Theta_s) = \frac{2}{\pi}V_B/(v_F^{K'}-v_F^{K})$. If $V_B\neq0$ and $v_F^{K}\approx v_F^{K'}$, one obtains $\Theta_s \approx \pi/4$, whereas if $V_B=0$, then $\Theta_s=0$.  The renormalized velocities become
\begin{align}
\Lambda_{I}^{\rm SLG}  =& R^{\rm SLG}_{11} \cos^{2} \Theta_s + R^{\rm SLG}_{22} \sin^{2} \Theta_s \nonumber\\
&~~~~~~~~~~~ - 2 R^{\rm SLG}_{12} \sin \Theta_s \cos \Theta_s,\\
\Lambda_{II}^{\rm SLG}  =& R^{\rm SLG}_{11} \sin^{2} \Theta_s + R^{\rm SLG}_{22} \cos^{2} \Theta_s\nonumber\\
&~~~~~~~~~~~ + 2 R^{\rm SLG}_{12} \sin \Theta_s \cos \Theta_s.
\end{align}
In Fig.~\ref{fig:v1v2} we show the renormalized velocities $\Lambda^{\rm SLG}_I$ and $\Lambda^{\rm SLG}_{II}$ as a function of the strength of the incident radiations,  as well as for two possible differences of the polarization angle ($\theta$): $\theta = \pi$ where the left and right halves
of the graphene layer are irradiated with left and right circularly polarized light and for $\theta = 0.8\pi$ where the polarization difference is slightly less.
Interestingly, we find  that the velocities   are strongly renormalized as a function of the amplitude of the light. They also depend on the difference in polarization of light impinging  on the two halves of the SLG.

We next compute the correlation functions, which are same for both $\sigma=\{\uparrow\downarrow\}$ spins, of the fermions as~\cite{fradkin,raosen}:
\begin{align}
\langle \Psi_{K \sigma } (y,t) {\Psi}^{\dagger}_{K \sigma } (0,0)\rangle &\sim \exp( \langle\Phi_{K \sigma } (y,t) {\Phi}^{\dagger}_{K \sigma } (0,0)\rangle)\\ \nonumber 
\Phi_{K \{\uparrow \downarrow\} } &= \frac{\Phi_{K c} \pm \Phi_{K s}}{\sqrt{2}} 
\end{align}
If $v_{i}$ is the velocity of $i^{th}$ mode, one can further write,
\begin{align}
\langle\Phi_{i}(y,t) \Phi_{j}(0,0)\rangle = -\frac{1}{4\pi}\text{ln}(y-v_{i}t)\delta_{ij},
\end{align}
which gives us,
\begin{align}
&\langle\Psi_{K \sigma } (y,t)  {\Psi}^{\dagger}_{K \sigma} (0,0)\rangle\nonumber\\ 	
&\sim \frac{1}{(y-\Lambda^{\rm SLG}_{I} t)^{ \frac{\cos^{2} \Theta_s}{2} }}
\frac{1}{(y-\Lambda^{\rm SLG}_{II} t)^{ \frac{\sin^{2} \Theta_s}{2} }}\frac{1}{(y-v^K_F t)^{ \frac12 }};\\
&\langle\Psi_{K' \sigma} (y,t)  {\Psi}^{\dagger}_{K' \sigma} (0,0)\rangle\nonumber\\ 	
&\sim \frac{1}{(y-\Lambda^{\rm SLG}_{I} t)^{ \frac{\sin^{2} \Theta_s}{2} }}
\frac{1}{(y-\Lambda^{\rm SLG}_{II} t)^{ \frac{\cos^{2} \Theta_s}{2} }}\frac{1}{(y-v^{K'}_F t)^{ \frac12 }}.
\end{align}
We obtain $\Theta_s\approx \pi/4$ (thus, $\sin^2\Theta_s\approx \cos^2\Theta_s\approx 1/\sqrt{2}$) for the relevant parameters, with weak dependence on the amplitude of the driving and  the polarization angle (not shown). It is easy to check that, if one turns off the interaction, the correlation functions become each of a fermionic mode with velocities $v_F^{K}$ or $v_F^{K'}$.

\subsection{Bi-layer graphene}
For the case of BLG, one can proceed similar to the SLG case. We start by writing the charge sector as 
\begin{align}
H^{c} & \equiv \sum_{\alpha } Q^{o}_{\alpha \alpha}  \nabla \Phi_{\alpha c} \nabla \Phi_{\alpha c}+ \sum_{\alpha \alpha' } Q_{\alpha \alpha'}  \nabla \Phi_{\alpha c} \nabla \Phi_{\alpha' c},
\end{align}
where 
\begin{align}
Q^{o}
=
\begin{bmatrix}
v^{1}_{F} & 0 & 0 & 0 \\
0 & v^{2}_{F} & 0 & 0 \\
0 & 0 & v^{1}_{F} & 0 \\
0 & 0 & 0 & v^{2}_{F} \\
\end{bmatrix};~
Q
=
\begin{bmatrix}
P & P \\
P & P
\end{bmatrix}; 
~P
=
\frac{1}{\pi}\begin{bmatrix}
V_{A} & V_{B} \\
V_{B} &  V_{A}
\end{bmatrix},
\end{align}
where $v_F^{1_K}\approx v_F^{1_{K'}} = v_F^1$ and $v_F^{2_K}\approx v_F^{2_{K'}} = v_F^2$. The form of the $Q$ and $P$ matrices arises from the computation of the scattering matrix elements. We proceed by performing another transformation -
\begin{align}
\Phi^{+}_{\eta c} = \frac{\Phi_{\eta_K c} + \Phi_{\eta_{K'} c}}{\sqrt{2}};~~ \Phi^{-}_{\eta c} = \frac{\Phi_{\eta_K c} - \Phi_{\eta_{K'} c}}{\sqrt{2}} ,
\end{align}
where $\eta =1,2$, to write
\begin{align}
\nonumber
H^{c}= \sum_{\eta} v^{\eta}_F [ (\nabla \Phi^{+}_{\eta c})^{2} + (\nabla \Phi^{-}_{\eta} )^{2}&]    \\ +\frac{2}{\pi} \sum_{\eta \eta' } V_{\eta \eta' \eta' \eta}&  ( \nabla \Phi^{+}_{\eta c} \nabla \Phi^{+}_{\eta' c} ), \\
H_{+}^{c}= \sum_{\eta} v^{\eta}_F [ (\nabla \Phi^{+}_{\eta c})^{2} ]    +\frac{2}{\pi} \sum_{\eta \eta' }& V_{\eta \eta' \eta' \eta}  ( \nabla \Phi^{+}_{\eta c} \nabla \Phi^{+}_{\eta' c} ), \\
H_{-}^{c}= \sum_{\eta} v^{\eta}_F [ (\nabla \Phi^{-}_{\eta c})^{2} ]  & ,
\end{align}
allowing us to further write the $(+)$ sector as 
\begin{align}
H^{+}_{c} =    & 
\begin{bmatrix}
	\nabla \Phi^{+}_{1 c} & \nabla \Phi^{+}_{2 c}
\end{bmatrix}
R^{\rm BLG}
 \begin{bmatrix} \nonumber
 \nabla \Phi^{+}_{1 c} \\
 \nabla \Phi^{+}_{2 c}
 \end{bmatrix} , \\ 
R^{\rm BLG} =& 
 \begin{bmatrix}
 v^{1}_{F}+\frac{2}{\pi}V_{A} & \frac{2}{\pi}V_{B} \\
 \frac{2}{\pi}V_{B} & v^{2}_{F} + \frac{2}{\pi}V_{A},
 \end{bmatrix} 
\end{align}
where it is evident that only the $(+)$ modes are renormalized. This sector can then be diagonalized using the canonical transformation
\begin{align}
\Phi^{+}_{1 c} = \cos\Theta_b \tilde{\Phi}_{1 c} + \sin\Theta_b \tilde{\Phi}_{2 c} ,\\ \nonumber
\Phi^{+}_{2 c} = -\sin\Theta_b \tilde{\Phi}_{1 c} + \cos\Theta_b \tilde{\Phi}_{2 c},	
\end{align} 
with $\tan(2\Theta_b) = \frac{4}{\pi}V_B/(v_F^2-v_F^1)$. If $V_B\neq0$ and $v_F^1\approx v_F^2$, one obtains $\Theta_b \approx \pi/4$, whereas if $V_B=0$, then $\Theta_b=0$.  The renormalized velocities of the $(+)$ sector become
\begin{align}
\Lambda^{\rm BLG}_{I}  =& R^{\rm BLG}_{11} \cos^{2} \Theta_b + R^{\rm BLG}_{22} \sin^{2} \Theta_b \nonumber\\
 &~~~~~~~~~~~ - 2 R^{\rm BLG}_{12} \sin \Theta_b \cos \Theta_b,\\
\Lambda^{\rm BLG}_{II}  =& R^{\rm BLG}_{11} \sin^{2} \Theta_b + R^{\rm BLG}_{22} \cos^{2} \Theta_b \nonumber \\
&~~~~~~~~~~~ + 2 R_{12} \sin \Theta_b \cos \Theta_b,
\end{align}
whereas, for the ($-$) sector, the modes remain unrenormalized. In Fig.~\ref{fig:v1v2} we show the renormalized velocities $\Lambda^{\rm BLG}_I$ and $\Lambda^{\rm BLG}_{II}$ as a function of the strength of the incident radiation, as well as for two values of the polarization angle ($\theta$), which show strong renormalization.

In terms of these fields, the fields of the original bosonic operator become
\begin{align}
&	\Phi_{1_K c} = \frac{1}{\sqrt{2}} [ \cos\Theta \tilde{\Phi}_{1 c} + \sin\Theta \tilde{\Phi}_{2 c} + \Phi^{-}_{1c} ]\nonumber \\
&\Phi_{2_K c} = \frac{1}{\sqrt{2}} [ -\sin\Theta \tilde{\Phi}_{1 c} + \cos\Theta \tilde{\Phi}_{2 c} + \Phi^{-}_{2c} ]\nonumber\\
&\Phi_{1_{K'} c} = \frac{1}{\sqrt{2}} [ \cos\Theta \tilde{\Phi}_{1 c} + \sin\Theta \tilde{\Phi}_{2 c} - \Phi^{-}_{1c} ]\nonumber\\
&\Phi_{2_{K'} c} = \frac{1}{\sqrt{2}} [ -\sin\Theta \tilde{\Phi}_{1 c} + \cos\Theta \tilde{\Phi}_{2 c} - \Phi^{-}_{2c} ]. \nonumber\\
&\Phi_{g \{\uparrow \downarrow\} } = \frac{\Phi_{g c} \pm \Phi_{g s}}{\sqrt{2}} ~;~ g = \{1_{K},1_{K'},2_{K},2_{K'}\} .
\end{align}
\begin{widetext}
Similar to the case of single-layer graphene, the correlation functions become:
\begin{align}
&\langle\Psi_{1_{(K,K')} } (y,t)  {\Psi}^{\dagger}_{1_{(K,K')} } (0,0)\rangle
\sim \frac{1}{(y-\Lambda^{\rm BLG}_{I} t)^{ \frac{\cos^{2} \Theta_b}{4} }}
\frac{1}{(y-\Lambda^{\rm BLG}_{II} t)^{ \frac{\sin^{2} \Theta_b}{4} }} 
\frac{1}{(y-v^1_{F} t)^\frac34}; \\
&\langle\Psi_{2_{(K,K')} } (y,t)  {\Psi}^{\dagger}_{2_{(K,K')} } (0,0)\rangle
\sim \frac{1}{(y-\Lambda^{\rm BLG}_{I} t)^{ \frac{\sin^{2} \Theta_b}{4} }}
\frac{1}{(y-\Lambda^{\rm BLG}_{II} t)^{ \frac{\cos^{2} \Theta_b}{4} }} 
\frac{1}{(y-v^2_{F} t)^\frac34};
\end{align}
\end{widetext}
We obtain $\Theta_b\approx \pi/4$ (thus, $\sin^2\Theta_b\approx \cos^2\Theta_b\approx 1/\sqrt{2}$) for the relevant parameters, with weak dependence on the amplitude of the driving and the polarization angle (not shown). Similar to the case of SLG. it is easy to check that, if one turns off the interaction, the correlation functions become each of a fermionic mode with velocities $v_1$ or $v_2$.

\section{Summary}
For  experimental realization, the crucial requirements are  that the topological mass gap, $m=\lambda^2\gamma/\omega$  be larger than the temperature scale and the driving frequency  be larger than the other energy scales. The intensity of the circularly polarized drive ($I=\frac12c\epsilon_0E^2$) can be written as $I\approx 10^{14} \alpha_d^2(\hbar\omega/t)$~W/cm$^2$, where the unitless parameter $\alpha_d=eAa_0/\hbar$ characterizes the driving amplitude. Assuming the topological mass to be of the order of meV and the driving frequency to be order of an electron-volt, one obtains $\alpha_d\sim 10^{-2}$, which in turn determines the required intensity of the drive. In an experimental set up, the possibility of heating may also need more careful consideration.

To summarize, we have studied the possibility of tunable chiral Luttinger liquid states at the interface of driven, topologically distinct states in two dimensions, specifically focusing on single and bi-layer graphene systems. The nature of the gap-opening of the bulk allows us to consider the effective interaction among the electrons at the topological steady-states to be effectively time-independent so that  we can apply standard bosonization techniques to  these interacting steady states. Our results suggest that these systems can act as a platform for highly tunable chiral Luttinger liquids, which can be further studied  experimentally.

%We neglected a number of possible issues in treating the system in simplistic manner, such as any heating effect, which may limit the scope of the applicability of the theory in two ways. First, the heating and the resulting phonon modes will essentially broaden the energy-levels and we need to have a large enough bulk-gap to distinguish the topological edge-modes from the bulk states.

\noindent {\it Acknowledgments} 

The research of AK was supported by funding from SERB, DST (Gov. of India), MHRD (Gov. of India) and DAE (Gov. of India). SB acknowledges support from UGC (Gov. of India). We also acknowledge HPC fecility of IIT Kanpur for computational work.

\renewcommand{\thefigure}{A\arabic{figure}}
\setcounter{figure}{0}
\renewcommand{\theequation}{A\arabic{equation}}
\setcounter{equation}{0}
\section*{Appendix}
In this appendix we write the Van Vleck expansion up to second order for the bi-layer graphene, which gives rise to a small difference between the velocities $v_1$ and $v_2$ of the topological edge-modes at $K$ and $K'$ momentum points. The second order i.e. $\mathcal{O}(\omega^{-2})$ correction to $H_{\rm eff}$, under the Van Vleck high-frequency expansion, is given by 
$$\frac{1}{2\omega^2}\displaystyle\sum_{n=1}^{\infty}\frac{1}{n^2}([[\mathcal{H}_n,\mathcal{H}_0],\mathcal{H}_{-n}]+ h.c.) $$
In our case, due to the sine and cosine nature of the driving, $\mathcal{H}_{\pm1}$ are the only non-zero $n\neq0$ Fourier coefficients of the Hamiltonian and the resulting correction term reduces to calculating $\frac{1}{2\omega^2}([[\mathcal{H}_1,\mathcal{H}_0],\mathcal{H}_{-1}])$. For convenience of notation, we introduce the following: $ \alpha= A_x +A_ye^{-i\theta} ,~~\beta = A_x-A_ye^{i\theta},~~\Gamma=A_x +A_ye^{i\theta},~~\delta= A_x-A_ye^{-i\theta}$. Further we use : $C\equiv \frac{\lambda^2 \gamma}{4\omega}\cos\theta$, $B\equiv 1-\frac{\lambda^2}{4\omega^2}(\alpha\Gamma+\beta\delta) $, $D\equiv \frac{\lambda^2}{2\omega^2}\delta\Gamma$, $E\equiv \frac{\lambda^2}{2\omega^2}\alpha\beta$ to obtain
\begin{widetext}
\begin{align}
\label{2ndordHam}
H_{\rm eff}= \begin{pmatrix}
-C & \nu \pi^{\dagger}B+E\nu\pi & 0 & t_p E \\
\nu \pi B+ D\nu\pi^{\dagger} & C & t_p B  & 0 \\
0 & t_p B &  -C  & \nu \pi^{\dagger}B+ E \nu\pi \\
t_p D & 0 & \nu \pi B+D\nu\pi^{\dagger} & C
\end{pmatrix}.
\end{align}
\end{widetext}
It is useful here to note the relations $B=1-\frac{\lambda^2(A_x^2+A_y^2)}{2\omega^2}$ and $D^{\dagger}=E$.  
%\left(1-\frac{\lambda^2}{4\omega^2}(\alpha\Gamma+\beta\delta)\right)
%\frac{\lambda^2}{2\omega^2}\alpha\beta
 %\frac{\lambda^2}{2\omega^2}\delta\Gamma
Using this Hamiltonian we proceed to calculate the effective $2$-band low energy sector. This can be done in similar manner of the main text~\cite{McCann2006} to have
\begin{align}
\label{2ndordlowenEffH}
H^L_{\rm eff}=\begin{pmatrix}
 (H^L_{\rm eff})_{1,1}& (H^L_{\rm eff})_{1,2}\\
 (H^L_{\rm eff})_{2,1}& (H^L_{\rm eff})_{2,2}
\end{pmatrix},
\end{align}
where the matrix elements are as follows
\begin{align}
(H^L_{\rm eff})_{1,1}&= -C-\frac{C\nu^2}{C^2+t_p^2 B^2}\times\nonumber \\
&((B^2+ED)\pi^{\dagger}\pi +BD(\pi^{\dagger})^2+ EB\pi^2), \\
(H^L_{\rm eff})_{1,2}&= t_p E - \frac{t_p B\nu^2}{C^2+t_p^2 B^2}\times\nonumber\\
& (B^2(\pi^{\dagger})^2+E^2\pi^2 + 2 EB \pi^{\dagger}\pi),
\end{align}

\begin{align}
(H^L_{\rm eff})_{2,1} &= t_p D-  \frac{t_p B\nu^2}{C^2+t_p^2 B^2}\times\nonumber\\
& (B^2\pi^2+ D^2(\pi^{\dagger})^2+ 2BD\pi^{\dagger}\pi), \\
(H^L_{\rm eff})_{2,2} &= C+\frac{C\nu^2}{C^2+t_p^2 B^2}\times\nonumber\\
& ((B^2+ED)\pi^{\dagger}\pi +BD(\pi^{\dagger})^2 + EB\pi^2),
\end{align}
which is valid in the low energy regime, when, $\epsilon\ll t_p\left(1- \frac{\lambda^2}{2\omega^2}(A_x^2+A_y^2)\right)$, which holds if one ensures that $\frac{\lambda^2A_xA_y}{\omega}\ll t_p\left(1- \frac{\lambda^2}{2\omega^2}(A_x^2+A_y^2)\right)$. One can make some simplifications to the above elements of $H^L_{\rm eff}$  by making approximations  where terms of $\mathcal{O}(\omega^{-4})$ are dropped out. These approximations are, $(B^2 +ED) \approx 1-\frac{\lambda^2}{2\omega^2}(A_x^2+A_y^2)$, $BD\approx D= \frac{\lambda^2}{2\omega^2}\delta\Gamma$, $EB\approx E =\frac{\lambda^2}{2\omega^2}\alpha\beta$ and $E^2=D^2=0$.  Additionally, given the condition for the low energy regime it follows that $t_p^2 B^2\gg C^2$. Using them one can re-examine the off -diagonal terms of $H^L_{\rm eff}$,
\begin{align}
(H^L_{\rm eff})_{1,2} &\approx t_pE - \frac{\nu^2}{t_p B}(B^2(\pi^{\dagger})^2+2E\pi^{\dagger}\pi)\\
(H^L_{\rm eff})_{2,1} &\approx t_p D - \frac{\nu^2}{t_p B}(B^2\pi^2+2D\pi^{\dagger}\pi).
\end{align}
One can compare the above off-diagonal terms to the off-diagonal terms for the low energy effective Hamiltonian computed in Eq. \eqref{eq:lowEnergyHam}  where only the $\mathcal{O}({\omega^{-1}})$ correction from the driving had been included. The modifications coming from the $t_pD$ and $t_pE$ kind of terms here, as higher order driving effects, are responsible for the observed asymmetry of the Fermi velocities of the chiral Luttinger edge modes in this system.  Thus these are significant in the regime that the edge modes are observed under the application of driving and are a manifestation of the long range hoppings induced by the drive. 

\end{document}